\documentclass[conference]{IEEEtran}
\IEEEoverridecommandlockouts
\usepackage{color}
\definecolor{red}{RGB}{255,0,0}
\usepackage{cite}
\usepackage{amsmath,amssymb,amsfonts}
\usepackage{algorithmic}
\usepackage{graphicx}
\usepackage{textcomp}
\usepackage{stfloats}
\usepackage{xcolor}
\usepackage[caption=false,font=footnotesize,labelfont=rm,textfont=rm]{subfig}
\usepackage{stfloats}
\usepackage{cuted}
\def\BibTeX{{\rm B\kern-.05em{\sc i\kern-.025em b}\kern-.08em
    T\kern-.1667em\lower.7ex\hbox{E}\kern-.125emX}}
\begin{document}

\title{Distortion-Aware Hybrid Beamforming for Integrated Sensing
	and Communication\\

\thanks{Z. Zhang, Y. Xiu, Z. Wu and N. Wei are with the National Key Laboratory of Wireless Communications, University of Electronic Science and Technology of China, Chengdu 611731, China (e-mail: zzycu@std.uestc.edu.cn;  xiuyue12345678@163.com; zixing\_wu2025@163.com; wn@uestc.edu.cn).
	
Phee Lep Yeoh is with the School of Science, Technology and Engineering,
University of the Sunshine Coast, Sunshine Coast, QLD 4556, Australia
(e-mail: pyeoh@usc.edu.au).

G. Liu is with the China Mobile Research Institute, Beijing 100053, China
(email: liuguangyi@chinamobile.com).

The corresponding author is Ning Wei.
 }
}

\author{Zeyuan Zhang,~Yue Xiu,~Phee~Lep Yeoh,~\IEEEmembership{Senior Member,~IEEE}, Guangyi Liu,\\Zixing Wu,
	~Ning Wei,~\IEEEmembership{Member,~IEEE}}

\maketitle

\begin{abstract}
This paper investigates a practical partially-connected hybrid beamforming transmitter for integrated sensing and communication (ISAC) with distortion from nonlinear power amplification. For this ISAC system, we formulate a communication rate and sensing mutual information maximization problem driven by our distortion-aware hybrid beamforming design. To address this non-convex  problem, we first solve for a fully digital beamforming matrix by  alternatively solving three sub-problems using manifold optimization (MO) and our derived closed-form solutions. The analog and digital beamforming matrices are then obtained through a decomposition algorithm. Numerical results demonstrate that the proposed algorithm can improve overall ISAC performance compared to traditional beamforming methods.
\end{abstract}
%
\section{Introduction}
In recent years, integrated sensing and communication (ISAC) has emerged as a promising technology for 6G and future networks \cite{1}. Compared to traditional separate radar and communication systems, ISAC simultaneously enables communication and radar sensing using a unified waveform and hardware platform. Through collective design, both spectral efficiency and sensing accuracy can be mutually enhanced with favorable hardware costs. Moreover, ISAC is compatible with prevailing wireless technologies such as reconfigurable intelligent surfaces (RIS), unmanned aerial vehicles (UAVs), and movable antennas (MAs), unlocking the potential for ubiquitous connectivity and unprecedented frontiers \cite{2,3,4}.

Motivated by the aforementioned advantages, numerous works have aimed at improving the sensing and communication performance of ISAC systems. In \cite{5}, the authors investigated a  hybrid beamforming design for ISAC  with the goal of minimizing the sensing Cramér-Rao bound (CRB) while satisfying the signal-to-interference-plus-noise ratio (SINR) constraints for individual communication users. In \cite{6}, the authors studied beamforming design with imperfect channel state information (CSI) in a cooperative ISAC system aided by movable antennas, where the problem is formulated to minimize the transmit power while ensuring that the target position estimation meets the CRB requirements and satisfies the worst-case communication rate constraint.

However, the performance of multi-antenna systems is limited by amplifier nonlinearities. For a real wireless transmitter, power amplifiers (PAs) typically operate close to their saturation point in order to be energy efficient, where their inherent nonlinearity inevitably causes signal distortion. In \cite{7}, the authors demonstrated how PA distortion affects the beam pattern in multiple-input-multiple-output (MIMO) transmitters. This model was further utilized to determine the distortion-aware power allocation, beamforming, and energy efficiency maximization problem in multi-user multi-input single-output (MU-MISO) systems \cite{8,7,10}. However, we note that existing ISAC-related references usually assume perfect linear amplification.

To address this gap, in this work, we consider a partially-connected ISAC transmitter performing sensing via communication resources while being distorted by PAs. We formulate a communication rate and sensing mutual information (MI) maximization problem driven by hybrid beamforming design. Specifically, we first solve for a fully digital beamforming matrix to address the highly non-convex problem with the help of manifold optimization, and then the analog and digital beamforming matrices are obtained through a decomposition algorithm.

The rest of the paper is organized as follows. In Section II, we introduce the ISAC system model under nonlinear amplification and formulate the optimization problem. In Section III, we solve the hybrid beamforming via an alternating optimization (AO) algorithm. In Section IV, we examine the efficacy of the proposed beamforming scheme through numerical examples. We conclude the paper in Section V.
\section{System Model and Problem Formulation}
\begin{figure}[t]
	\centering
	\includegraphics[width=3in]{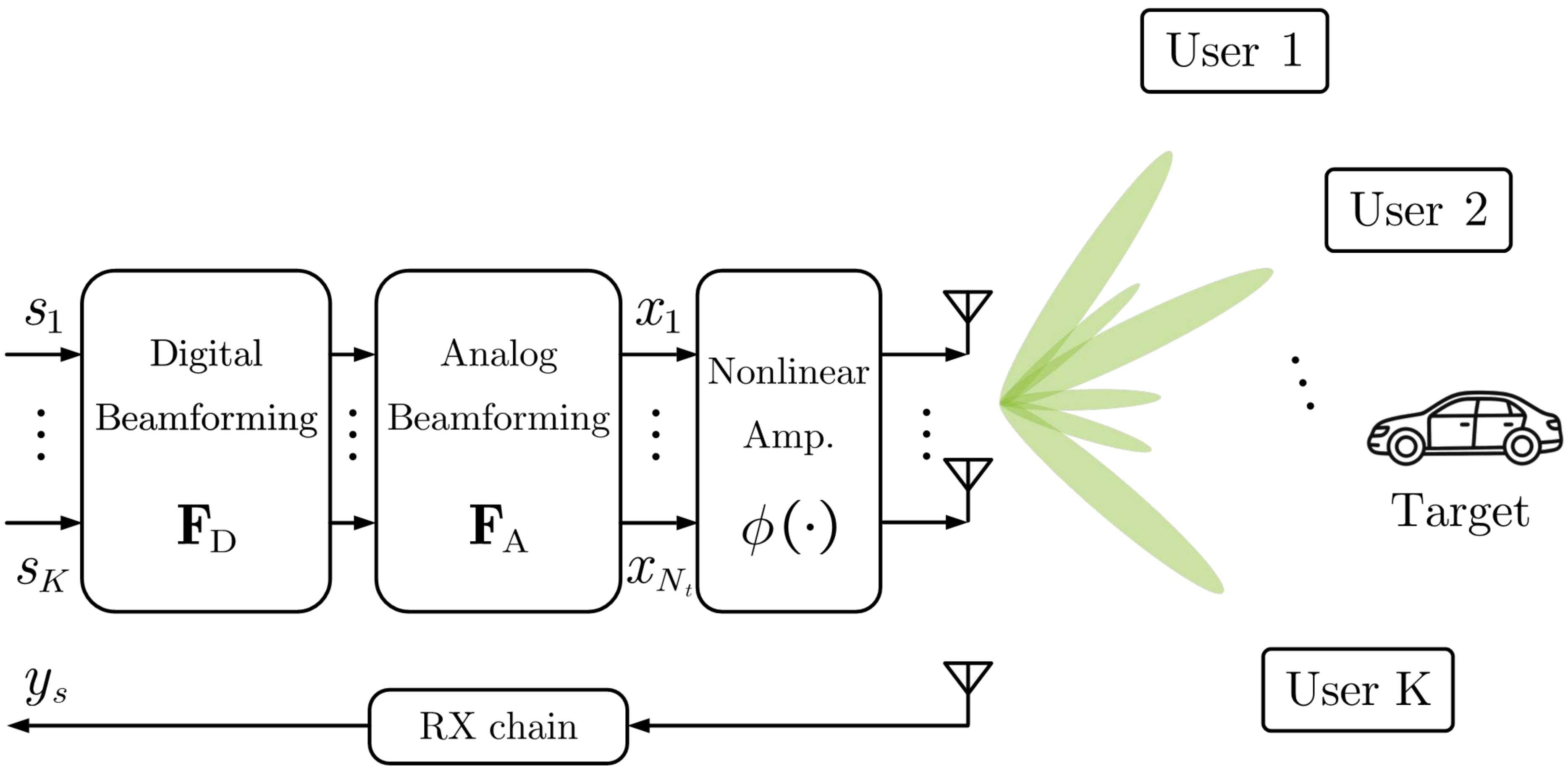}
	\caption{System model of the hybrid beamforming ISAC with nonlinear amplification.}
	\label{f1}
\end{figure}
As illustrated in Fig.~\ref{f1}, we consider a monostatic ISAC base station (BS)  simultaneously serves $K$ users and senses a single target. Each user, as well as the sensing receiver (SR), is equipped with a single antenna. In the considered system, the BS employs a partially-connected uniform linear array (ULA) with $N_t$ antennas, which is composed of $N_{\text{RF}}$ non-overlapping subarrays each driven by a dedicated  transmit chain. Without loss of generality, we assume the number of data streams is equal to the number of users. Under this configuration, the BS can sense the target via the communication signals without increasing the number of transmit chains.
\subsection{Channel Model}
We consider a ULA with $\frac{\lambda}{2}$ antenna spacing, where the field response vector (FRV) of the $l$-th path of the $k$-th user is given by
\begin{equation}\mathbf{a}_{k,l}=\frac{1}{\sqrt{N_t}}\begin{bmatrix}1,e^{j\pi \cos\theta_{k,l}},\cdots,e^{j\pi\left( N_t-1\right) \cos\theta_{k,l}}\end{bmatrix}^T\in\mathbb{C}^{N_t\times1},\end{equation}
where $\theta_{k,l}$ denotes the angle of departure (AoD) of the $l$-th path of user $k$. Accordingly, the communication channel between the BS and user $k$ can be expressed as
\begin{equation}\mathbf{h}_k=\sqrt{\frac{N_t}{L}}\sum\nolimits_{l=1}^{L}\rho_{k,l}\mathbf{a}_{k,l},\end{equation} %
where $ L $ denotes the number of multipath components, and $ \rho_{k,l} $ is the complex channel gain of the $ l $-th path of user $ k $. For proof of concept, we consider a single point target case with single path model. With $ \theta_s $ being the sensing target's azimuth angle, the FRV of the sensing target is given by
\begin{equation}\mathbf{a}_{s}=\frac{1}{\sqrt{N_t}}\begin{bmatrix}1,e^{j\pi\cos\theta_s},\cdots,e^{j\pi\left( N_t-1\right) \cos\theta_s}\end{bmatrix}^T\in\mathbb{C}^{N_t\times1}.\end{equation} %
\subsection{Signal Model with Nonlinear Amplification}
Let $\mathbf{s}=[s_1,\cdots,s_K]^T$ be the $ K $ streams transmit signal satisfying  $\mathbb{E}\{{\mathbf{ss}^H}\}=\mathbf{I}_K$, the signal vector at the input of the the PAs stage is given by
\begin{equation}\mathbf{x}=\mathbf{F}_\mathrm{A}\mathbf{F}_\mathrm{D}\mathbf{s},\end{equation} %
where $\mathbf{F}_{\mathrm{A}}=\mathrm{blkdiag}\left( \mathbf{f}^{\mathrm{A}}_{1},\mathbf{f}^{\mathrm{A}}_{2},\cdots,\mathbf{f}^{\mathrm{A}}_{N_{RF}}\right) \in\mathbb{C}^{N_{t}\times N_{RF}}$ and  $\mathbf{F}_{\mathrm{D}}=[\mathbf{f}^{\mathrm{D}}_{1},\cdots,\mathbf{f}^{\mathrm{D}}_{K}]\in\mathbb{C}^{N_{RF}\times K}$ are the the analog and digital beamforming matrices, respectively. $ N_{RF} $ is the number of transmit chains. For simplicity, we denote  $\mathbf{F}=\mathbf{F}_{\mathrm{A}}\mathbf{F}_{\mathrm{D}}$ as the overall beamforming matrix.  Since the partially-connected structure is employed, $\mathbf{F}_{\mathrm{A}}$ is a block diagonal matrix with  $|\mathbf{f}^{\mathrm{A}}_i(j)|=1,j=1,\cdots,N_{t}$. According to the Bussgang's theorem, the nonlinear amplification process, for a given linear input $ \mathbf{x} $, can be expressed as
\begin{equation}\phi(\mathbf{x})=\mathbf{B}\mathbf{x}+\mathbf{e},\end{equation} %
where $\mathbf{B}\in\mathbb{C}^{N_t\times N_t} $ is a diagonal matrix whose entries along the diagonal are given by $[\mathbf{B}]_{ii}=\mathbb{E}[\phi(x_{i})x_{i}^{*}]/\mathbb{E}[|x_{i}|^{2}]$ and $ \mathbf{e} $ is the distortion component satisfying $\mathbb{E}[\mathbf{xe}^H]=\mathbf{0}$. In this paper, we consider a third-order polynomial to model the PA nonlinearity. As a consequence, we have
\begin{equation}\mathbf{B}=\beta_1\mathbf{I}_{N_t}+2\beta_3\mathbf{FF}^H\odot\mathbf{I}_{N_t},\end{equation} %
where $ \beta_1 $, $ \beta_3 \in \mathbb{C}$ are the model parameters. It can also be shown that the covariance matrix of $ \mathbf{e} $ is given by \cite{7}
\begin{equation}\mathbf{C}_{\mathbf{e}}=2{\left|\beta_{3}\right|}^{2}\mathbf{F}\mathbf{F}^{H}\odot| \mathbf{F}\mathbf{F}^{H}|^2 .\end{equation}%
After nonlinear amplification, the distorted signals $ \phi\left( \mathbf{x}\right) \in\mathbb{C}^{N_{t}\times 1} $ are then spatially combined through the channel. For user $k  $, the received signal can be written as
%
\begin{eqnarray}
y_{k}={\mathbf{h}^H_k \mathbf{B}\mathbf{F}_\mathrm{A}\mathbf{f}^{\mathrm{D}}_ks_k}+{\mathbf{h}^H_k \mathbf{B}\mathbf{F}_\mathrm{A}\sum_{k^{\prime}\neq k}\mathbf{f}^{\mathrm{D}}_{k^{\prime}}s_{k^{\prime}}}+{\mathbf{h}^H_k \mathbf{e}}+{n_k},
\end{eqnarray}%
where $n_{k}\sim\mathcal{CN}(0,\sigma^{2}_{k})$ is the additive white Gaussian noise (AWGN) of  user $ k $. The signal-to-interference-and-noise-and-distortion ratio of user $k$ can be given by
\begin{equation}\mathrm{\gamma}_k=\frac{|\mathbf{h}_k^H\mathbf{B}\mathbf{F}_\mathrm{A}\mathbf{f}^{\mathrm{D}}_k|^2}{\sum_{k^{'} \neq k}|\mathbf{h}_k^H\mathbf{B}\mathbf{F}_\mathrm{A}\mathbf{f}^{\mathrm{D}}_{k^{'}}|^2+\mathbf{h}_k^H\mathbf{C}_\mathbf{e}\mathbf{h}_k+\sigma_k^2}.\end{equation} %
For the sensing link, the echo signal received by the SR is
%
\begin{equation}
			y_s={\alpha_s\mathbf{a}_s^H\mathbf{B}\mathbf{F}_{\mathrm{A}}\mathbf{F}_{\mathrm{D}}\mathbf{s}}+{\alpha_s\mathbf{a}_s^H\mathbf{e}}+{n_s},
\end{equation} %
where $ \alpha_s $ is the complex coefficients including the radar cross section (RCS) of the target, and $n_s\sim\mathcal{CN}(0,\sigma_s^2)$ is the AWGN for the sensing link. The signal-to-noise-and-distortion ratio of the sensing link is thus given by
\begin{equation}\mathrm{\gamma}_s=\frac{||\alpha_s\mathbf{a}_s^H\mathbf{B}\mathbf{F}_\mathrm{A}\mathbf{F}_\mathrm{D}||^2}{|\alpha_s|^2\mathbf{a}_s^H\mathbf{C}_\mathbf{e}\mathbf{a}_s+\sigma_s^2}.\end{equation} 
\newcounter{TempEqCnt}
\setcounter{TempEqCnt}{\value{equation}} 
\setcounter{equation}{16}                      
\begin{figure*}[hb]
		\hrulefill
	\centering
	\begin{subequations}
	\begin{alignat}{2}
		(\mathcal{P}3)~~~
		\max_{\mathbf{F},\mathbf{U},\mathbf{V}} \quad &\tilde{\mathcal{G}_1}\left( \mathbf{F},\mathbf{U},\mathbf{V}\right)= \varpi_c\sum_{k=1}^{K}\log_2\left(1+ {\frac{S_{k}}{I_{k}+D_k+\sigma_k^2}}\right)+ \varpi_s \log_2{\left(1+\frac{S_s}{D_s+\sigma_s^2} \right) } \label{pri}& \\
		&+\lambda_1		\left\| \mathbf{U}-\mathbf{F}\mathbf{F}^H\odot\mathbf{F}^*\mathbf{F}^T\right\|_F^2+\lambda_2\left\| \mathbf{V}-\mathbf{U}\odot\mathbf{F}\mathbf{F}^H\right\|_F^2, \nonumber\\
		\mbox{s.t.}\quad
		&{\left| \beta_{1}\right| ^{2}\operatorname{Tr}(\mathbf{\mathbf{F}\mathbf{F}}^H)+4\Re\left\lbrace\beta^*_{1}\beta_{3} \right\rbrace \operatorname{Tr}(\mathbf{U})+6|\beta_{3}|^{2}\operatorname{Tr}(\mathbf{V})= P_{tot}}.\label{pri2}
	\end{alignat}
\end{subequations}
\end{figure*}
 \setcounter{equation}{\value{TempEqCnt}}
\subsection{Performance Metric and Problem Formulation}
To analyze the performance of the considered ISAC  system, we first characterize the communication performance with the achievable data rate. The achievable data rate of user $ k $ is 
 \begin{equation}R_k=\log_2\left( 1+\gamma_{k}\right) .\end{equation} %
For the sensing link, the conditional sensing mutual information is adopted as the performance metric, which is expressed as \cite{4}
 \begin{equation}R_s=\log_2\left( 1+\gamma_{s}\right) .\end{equation} %
Accordingly, to maximize the sum of achievable data rate and sensing mutual information, the optimization problem is formulated as
\begin{subequations}
	\begin{align}
		(\mathcal{P}0)~~~\max_{\mathbf{F}_{\mathrm{A}},\mathbf{F}_{\mathrm{D}}}&~{\mathcal{G}}\left( \mathbf{F}_{\mathrm{A}},\mathbf{F}_{\mathrm{D}}\right) \triangleq \varpi_c\sum_{k=1}^{K}R_k+\varpi_s R_s,\label{14a}\\
		\mbox{s.t.}~
		&\mathbf{F}_{\mathrm{A}}\in\mathcal{A},&\label{14b}\\
		&\mathbb{E}\left\lbrace \left\|\phi(\mathbf{F}_{\mathrm{A}}\mathbf{F}_{\mathrm{D}}\mathbf{s})\right\|^2\right\rbrace =P_{tot},&\label{14c}
	\end{align}
\end{subequations}%
where $ \varpi_c $ and $ \varpi_s $ are the communication and sensing weighting factors of the system, respectively, satisfying $ \varpi_c + \varpi_s =1 $. Constraint (\ref{14b}) corresponds to the partially-connected structure, i.e., the $ \mathbf{F}_{\mathrm{A}} $ belongs to a set of block matrices $ \mathcal{A} $, where each block is a $N_t/N_{RF} $ dimension vector with unit modulus elements. Constraint (\ref{14c}) is the overall power budget.
\section{Proposed Distortion-Aware ISAC Beamforming with Nonlinear Amplification}
The objective function and the constraint of the original problem $ \mathcal{P}0 $ are both highly nonconvex w.r.t. the beamforming matrices $\mathbf{F}_{\mathrm{A}}$ and $\mathbf{F}_{\mathrm{D}}$, which makes it challenging to solve. In this section, we propose a two-stage distortion-aware hybrid beamforming algorithm for the considered ISAC system. Due to the mutually coupled variables, we choose to determine a full digital beamforming matrix  $\mathbf{F}$ and then decompose it into digital and analog components. Accordingly, $\mathcal{P}_0$ can be sequentially fragmented into the following two problems:
\begin{subequations}
	\begin{align}
		(\mathcal{P}1)~~~\max_{\mathbf{F}}&~{\mathcal{G}_1}\left( \mathbf{F}\right) \triangleq \varpi_c\sum_{k=1}^{K}R_k+\varpi_s R_s,\label{16a}\\
		\mbox{s.t.}~
		&\text{(\ref{14c})}&\label{16b}
	\end{align}
\end{subequations}%
\begin{subequations}
\begin{align}
	(\mathcal{P}2)~~~\min_{\mathbf{F}_{\mathrm{A}},\mathbf{F}_{\mathrm{D}}}&~ {\mathcal{G}_2}\left( \mathbf{F}_{\mathrm{A}},\mathbf{F}_{\mathrm{D}}\right) \triangleq\left\|\mathbf{F}-\mathbf{F}_{\mathrm{A}}\mathbf{F}_{\mathrm{D}} \right\|_F^2,&\label{17a}  \\
	\mbox{s.t.}\quad
	&\text{(\ref{14b})}. &\label{17b}
\end{align}
\end{subequations}%
To reduce the order of the variable in $ \mathcal{P}1 $, we adopt the penalty function method and introduce two auxiliary variables $\mathbf{U}$ and $\mathbf{V}$. With proper  reformulation, $\mathcal{P}_1$ is recast into  $\mathcal{P}_3$, as shown at the bottom of this page, where $\lambda_1, \lambda_2 < 0$ are penalty factors. Furthermore, the variables involved in $\mathcal{P}_3$ are defined as follows:
\setcounter{equation}{17} 
\begin{equation}
	\begin{aligned}
	S_{k}&=\left| \mathbf{h}_k^H\left(\beta_1\mathbf{I}_{N_t}+2\beta_3\mathbf{FF}^H\odot\mathbf{I}_{N_t}\right) \mathbf{f}_k\right|^2,\\
	I_k&=\sum\nolimits_{i \neq k}\left| \mathbf{h}_k^H\left(\beta_1\mathbf{I}_{N_t}+2\beta_3\mathbf{FF}^H\odot\mathbf{I}_{N_t}\right) \mathbf{f}_i\right|^2,\\
 	S_s&=\left\|\alpha_s \mathbf{a}_s^H\left(\beta_1\mathbf{I}_{N_t}+2\beta_3\mathbf{FF}^H\odot\mathbf{I}_{N_t}\right) \mathbf{F}\right\|^2, \\
	D_k&=2{\left|\beta_{3}\right|}^{2}\mathbf{h}_k^H\left( \mathbf{F}\mathbf{F}^H\odot| \mathbf{F}\mathbf{F}^{H}|^2\right) \mathbf{h}_k,\\
	D_s&=2{\left|\beta_{3}\right|}^{2}{\left|\alpha_s\right|}^{2}\mathbf{a}_s^H\left( \mathbf{F}\mathbf{F}^H\odot| \mathbf{F}\mathbf{F}^{H}|^2\right) \mathbf{a}_s.
	\end{aligned}
\end{equation}%
In order to solve $\mathcal{P}_3$, we adopt an AO framework, where each variable is updated iteratively while keeping the others fixed at their values from the previous iteration. To be specific, $\mathcal{P}_3$ is decomposed into the following three subproblems:
\begin{itemize}
	\item Update the full digital beamforming matrix $ \mathbf{F} $:
\end{itemize}
\begin{alignat}{2}
	(\mathcal{SP}3.1)~~~
	\max_{\mathbf{F}} \quad &\tilde{\mathcal{G}}_1\left( \mathbf{F}\,|\,\mathbf{U}^{(t)},\mathbf{V}^{(t)}\right),   \\
	\mbox{s.t.}\quad
	&\text{(\ref{pri2})}.\nonumber
\end{alignat}
\begin{itemize}
	\item Update the auxiliary variable $ \mathbf{U} $:
\end{itemize}
\begin{alignat}{2}
(\mathcal{SP}3.2)~~~
\max_{\mathbf{U}} \quad &\tilde{{\mathcal{G}}_1}\left(\mathbf{U}\,|\, \mathbf{F}^{(t)},\mathbf{V}^{(t)}\right),\\
\mbox{s.t.}\quad
&\text{(\ref{pri2})}.\nonumber
\end{alignat}
\begin{itemize}
	\item Update the auxiliary variable $ \mathbf{V} $:
\end{itemize}
\begin{alignat}{2}
	(\mathcal{SP}3.3)~~~
	\max_{\mathbf{V}} \quad &\tilde{{\mathcal{G}}_1}\left(\mathbf{V}\,|\, \mathbf{F}^{(t)},\mathbf{U}^{(t)}\right),\\
	\mbox{s.t.}\quad
	&\text{(\ref{pri2})}.\nonumber
\end{alignat} %
where $ \mathbf{F}^{(t)} $, $ \mathbf{U}^{(t)}  $ and $ \mathbf{V}^{(t)}  $ denotes the value of $ \mathbf{F}  $, $ \mathbf{U} $ and $ \mathbf{V}  $ from the previous iteration. For notational simplicity,  the superscript $ (t) $ is omitted in the following.
Focusing on $ \mathcal{SP}3.1 $, we notice the objective function is still nonconvex. Though the gradient ascent (GA) method can be utilized to find a local stationary point, constraint (\ref{pri2}) is not necessarily fulfilled. To address this challenge, we propose a manifold optimization (MO) algorithm to deal with $ \mathcal{SP}3.1 $. Note the set of the complex matrix that fulfills the constraint of constant Frobenius norm corresponds to a Riemannian manifold $ \mathcal{M}\subseteq \mathbb{C}^{N_t\times K} $. The differentiability of a Riemannian manifold is well-defined, allowing us to reformulate $ \mathcal{SP}3.1 $ as an unconstrained problem limited on $ \mathcal{M} $. Specifically, the Riemannian gradient at a given point $ \mathbf{F}_0 $ on the manifold $ \mathcal{M} $, i.e., $ \mathrm{grad}~\tilde{\mathcal{G}}_1\left( \mathbf{F}_0\,|\,\mathbf{U},\mathbf{V}\right)  $, is given by the orthogonal projection of the Euclidean gradient $ \nabla\tilde{\mathcal{G}}_1\left( \mathbf{F}_0\,|\,\mathbf{U},\mathbf{V}\right) $ onto the tangent space $\mathcal{T}_{\mathbf{F}_0}\mathcal{M}=\{\mathbf{G}\in\mathbb{C}^{m\times n}\mid\Re\left\lbrace \langle\mathbf{G},\mathbf{F}_0\rangle_F\right\rbrace =0\}$. Thus, in the $ i $-th iteration, the update direction $ \mathbf{L} $ and the full digital beamforming  matrix $ \mathbf{F} $  can be determined by
	\begin{subequations}
\begin{equation}
		\mathbf{L}_{i}=\mathrm{grad}~\tilde{\mathcal{G}}_1\left( \mathbf{F}_i\,|\,\mathbf{U},\mathbf{V}\right)+r_i\mathrm{proj}_{\mathcal{T}_{\mathbf{F}_i}\mathcal{M}}(\mathbf{L}_{i-1}),
\end{equation}%
\begin{equation}
\mathbf{F}_{i+1}=\mathrm{retr}_{\mathbf{F}_i}(t_i\mathbf{L}_i),
\end{equation}\label{222}%
\end{subequations}%
where $ r_i=\frac{\|\mathrm{grad}~\tilde{\mathcal{G}}_1\left( \mathbf{F}_i\,|\,\mathbf{U},\mathbf{V}\right)\|_2^2}{\|\mathrm{grad}~\tilde{\mathcal{G}}_1\left( \mathbf{F}_{i-1}\,|\,\mathbf{U},\mathbf{V}\right)\|_2^2} $ is the Fletcher-Reeves parameter, $ t_i $ is the step size determined by the Armijo line search. The retraction $ \mathrm{retr}_{\mathbf{F}}\left( \cdot\right): \mathcal{T}_{\mathbf{F}}\mathcal{M}\rightarrow\mathcal{M}$ that maps the updated point back to the manifold itself is given by
\begin{equation}\begin{aligned}\operatorname{retr}_{\mathbf{F}}(t_i\mathbf{L}_i)=\frac{\sqrt{c_1}\left( \mathbf{F}+t_i\mathbf{L}_i\right) }{\left\| \mathbf{F}+t_i\mathbf{L}_i\right\|_F },\end{aligned}\end{equation} %
where $ c_1= \frac{P_{tot}-4\Re\left\lbrace\beta^*_{1}\beta_{3} \right\rbrace \operatorname{Tr}(\mathbf{U})-6|\beta_{3}|^{2}\operatorname{Tr}(\mathbf{V})}{\left| \beta_{1}\right|^2 }$ is the square of the desired Frobenius norm. Equations (\ref{222}) are repeated until $ \mathbf{F} $ converges.
The Euclidean gradient of the cost function $ \nabla\tilde{\mathcal{G}}_1\left( \mathbf{F}\,|\,\mathbf{U},\mathbf{V}\right)  $ in $ \mathcal{SP}3.1 $ is given in the \textbf{Appendix}.

With $\mathbf{F}$ obtained from the proposed MO algorithm, we focus on the remaining subproblems to update the auxiliary variables $ \mathbf{U} $ and $ \mathbf{V} $. Specifically, with $ \mathbf{F} $ and $ \mathbf{V} $ being fixed, $ \mathcal{SP}3.2 $ is a linear constrained quadratic problem. The Lagrangian of $ \mathcal{SP}3.2 $ can be derived as
	\begin{subequations}
	\begin{alignat}{2}
	\mathcal{L}\left( \mathbf{U},\mu\right)&=\, \mathrm{const} -\lambda_1\left\| \mathbf{U}-\mathbf{F}\mathbf{F}^H\odot\mathbf{F}^*\mathbf{F}^T\right\|_F^2\\&-\lambda_2\left\| \mathbf{V}-\mathbf{U}\odot\mathbf{F}\mathbf{F}^H\right\|_F^2 \nonumber
		+ \Re\left\lbrace\mu\left(  \operatorname{Tr}\left( \mathbf{U}\right)-c_2\right)   \right\rbrace ,
	\end{alignat}
\end{subequations} %
where $ 	c_2=\frac{P_{tot}-\left| \beta_{1}\right| ^{2}\left\| \mathbf{F}\right\| _F^2-6|\beta_{3}|^{2}\operatorname{Tr}(\mathbf{V})}{4\Re\left\lbrace\beta^*_{1}\beta_{3} \right\rbrace}.
 $
By applying  the Karush-Kuhn-Tucker (KKT) conditions, the closed-form solution of $ \mathbf{U} $ and dual variable $ {\mu}^\star $ can be derived as follows:
\begin{subequations}
\begin{equation}
\begin{aligned}
	\mathbf{U}^\star =
	\left( \frac{\mu^*}{2}\mathbf{I}_{N_t}+\lambda_1 \left|\mathbf{C_x} \right|^2 + \lambda_2 \mathbf{V} \odot \mathbf{C_x}^* \right)
	\oslash
	\mathbf{\Xi} ,
\end{aligned}
\end{equation}%
\begin{equation}
{\mu}^{\star} =
\frac{
	c_2^*-\operatorname{Tr} \left(
	\left( \lambda_1 \left|\mathbf{C_x} \right|^2 + \lambda_2 \mathbf{V}^* \odot \mathbf{C_x} \right)
	\oslash
\mathbf{\Xi} 
	\right)
}{
	\frac{1}{2} 
	\operatorname{Tr} \left(
	\mathbf{I}_{N_t} \oslash \mathbf{\Xi} 
	\right)
},
\end{equation}%
\end{subequations}%
where $\mathbf{\Xi}=\lambda_1\mathbf{1}_{N_t} + \lambda_2 \left|\mathbf{C_x} \right|^2 $, $ \mathbf{C_x}=\mathbf{F}\mathbf{F}^H $,   $ \mathbf{1}_{N_t} \in \mathbb{C}^{N_t\times N_t}$ is an all-one matrix, and $ \oslash $ denotes the element-wise division.
With $ \mathbf{F} $ and $ \mathbf{U} $ being fixed, $ \mathcal{SP}3.3 $ is a  least square problem with affine constraint. The optimal solution is the orthogonal projection of $\mathbf{U}\odot\mathbf{C_x}$ onto the hyperplane specified by $ \operatorname{Tr}\left(\mathbf{V} \right)= c_3  $, where
$ c_3=\frac{P_{tot}-\left| \beta_{1}\right| ^{2}\left\| \mathbf{F}\right\| _F^2-4\Re\left\lbrace\beta^*_{1}\beta_{3} \right\rbrace \operatorname{Tr}(\mathbf{U})}{6\left| \beta_{3}\right| ^2} $. Hence, the closed-form  solution of $ \mathcal{SP}3.3 $ is
 
\begin{equation}
	\mathbf{V}^{\star}=\mathbf{U}\odot\mathbf{C_x}-\frac{\operatorname{Tr}(\mathbf{U}\odot\mathbf{C_x})-c_3}{N_t}\cdot\mathbf{I}_{N_t}.
\end{equation}
The update of $ \mathbf{F} $, $ \mathbf{U} $ and $ \mathbf{V} $ stops when the objective function of $ \mathcal{P}3 $ is fully converged. 

With fixed full digital beamforming matrix $ \mathbf{F} $, the analog and digital beamforming matrices, $ \mathbf{F}_{\mathrm{A}}$ and $\mathbf{F}_{\mathrm{D}} $, can be obtained by  solving  $ \mathcal{P}2 $. This turns out to be a well-studied matrix decomposition problem, which can be solved by, e.g., orthogonal matching pursuit (OMP) or closed-form expression in a similar AO algorithm \cite{12}. In this paper, we adopt the second approach to obtain the final solution of $ \mathbf{F}_{\mathrm{A}}$ and $\mathbf{F}_{\mathrm{D}} $.
%
\section{Numerical Results}\label{IV}
In this section, we evaluate the performance of the proposed AO-based beamforming algorithm through numerical experiments in terms of the sum of the  communication rate and sensing MI (hereafter referred to as sum rate), convergence, and beam pattern. In the experiment, the sensing target is assumed to be located at $ \theta_s=60^\circ $, while the user locations are generated through random channel realization with $ \rho_{k,l} \sim \mathcal{CN}(0,1)$ and $ \theta_{k,l} \sim \mathcal{U}(0,\pi)$. Users and the target are considered to have the same noise level and weighting factors, i.e., $ \sigma_k^2=\sigma_s^2=N_0 $, $ \varpi_c=\varpi_s=0.5 $. The signal-to-noise ratio (SNR) is given by $ P_{tot}/N_0 $. Furthermore, we set $ \beta_{1}=1.14-0.08j $, $ \beta_{3}=-0.08 + 0.1j $, which is extracted  from the polynomial model of mm-wave AWMF-0108 PA \cite{13}.
\begin{figure}[!t]
	\centering
	\hspace{-0.15cm}\subfloat[]{
		\includegraphics[width=1.40in]{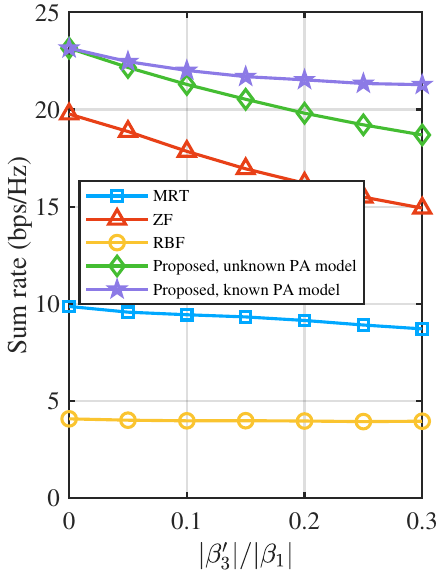}}\hspace{0.01cm}\hspace{0.2cm}
	\subfloat[]{
		\includegraphics[width=1.35in]{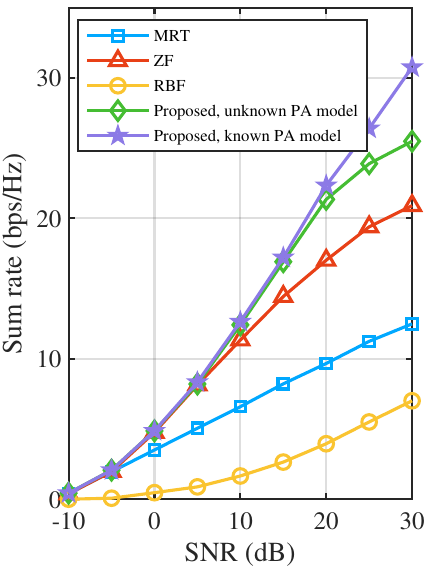}}
	\caption{Ergodic sum rate performance with  $ N_t=64 $, $ N_{RF}=16 $, $ K=2 $, $ L=5 $ and $ P_{tot}=13 \,$dBm. (a) Ergodic sum rate v.s. $ |\beta_3^{\prime}|/|\beta_1| $ with $ \text{SNR}=20 \,$dB. (b) Ergodic sum rate v.s. $ \text{SNR} $ with $ \beta_{1}=1.14-0.08j $ and $ \beta_{3}=-0.08 + 0.1j $.}
\label{f2}
\end{figure} 
\begin{figure}[t]
	\centering
	\includegraphics[width=2.35in]{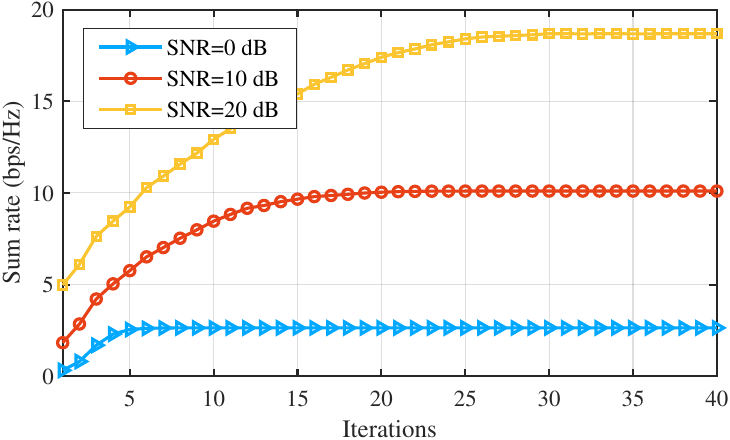}
	\caption{Average convergence behavior of the proposed MO algorithm while solving for the full digital beamforming matrix $ \mathbf{F} $. $ N_t=16 $, $ K=2 $, $ L=3 $, $ P_{tot}=13 $ dBm.}
	\label{f3}
\end{figure}

In Fig.~\ref{f2}, we investigate the ISAC  performance under different level of PA nonlinearity and SNR. The relevant simulation parameters are specified in the caption. Specifically, the proposed scheme with unknown PA model is obtained with the assumption of ideal amplification, while the proposed scheme with known PA model corresponds to the precoder designed by the complete algorithm with prior knowledge of the PA parameters. With $ 10^3 $ channel realizations, we compare the ergodic sum rate of the proposed algorithm and classical precoders including random beamforming (RBF), maximal-ratio transmission (MRT), and zero-forcing (ZF). In  Fig.~\ref{f2}(a), we alter the distortion strength by scaling $ \beta_3 $ to $ \beta_3^{\prime} $ while keeping $ \beta_1 $ unchanged.  As can be observed, except for the RBF precoder, the performance of all precoders tends to degrade evidently with the increase of nonlinear level $ |\beta_3^{\prime}|/|\beta_1| $. The proposed algorithm with known PA model outperforms the other algorithms, and the performance gap further increases as the nonlinear level $ |\beta_3^{\prime}|/|\beta_1| $  becomes larger. This is because the other algorithms do not account for the nonlinearity of the PAs, in other words, they are not distortion-aware. Fig.~\ref{f2}(b) further indicates that the gain of the proposed algorithm increases  with higher SNR. This is because the distortion is more significant  under low noise conditions.

In Fig.~\ref{f3}, we show the average convergence behavior while solving for the full digital beamforming matrix $ \mathbf{F} $, i.e., the sum rate versus the iterations of the proposed MO algorithm when $ \text{SNR}=\{0,10,20\} \,$dB. The relevant simulation parameters are given in the caption. We note that the convergence of the algorithm is faster at lower SNRs, because the distortion has a more prominent impact on the sum rate at high SNRs. Furthermore, since the proposed MO  algorithm is essentially similar to the conjugate gradient (CG) method, the convergence of this algorithm is guaranteed.

\begin{figure}[t]
	\centering
	\includegraphics[width=2.5in]{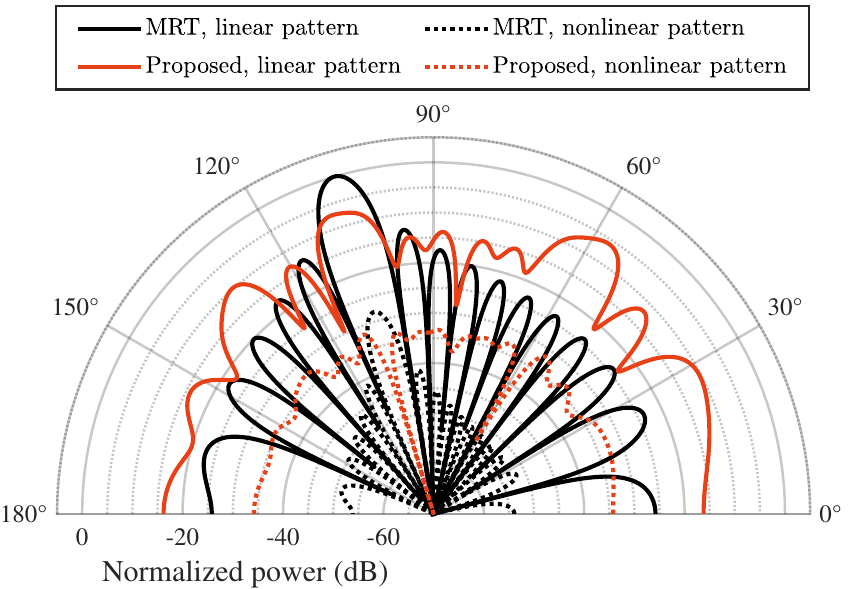}
	\caption{Linear and nonlinear beam pattern of the  proposed algorithm with known PA model and MRT. $ N_t=16 $, $ N_{RF}=4 $, $ K=1 $, $ L=1 $, $ P_{tot}=20 \,$dBm and $ \text{SNR}=25\, $dB.}
	\label{f4}
\end{figure}

Fig.~\ref{f4} shows the linear and nonlinear beam pattern of the proposed algorithm with known PA model, in comparison to the MRT precoder. The relevant simulation parameters are specified in the caption. Specifically, a single user is located at $\theta_1= 106^\circ$. With the MRT scheme, it is observed that the nonlinear pattern is also beamformed towards the user, as indicated in the black dot line. The proposed beamforming algorithm performs $ 7 \,$dB weaker in linear power at the user direction, which is a result of non-zero sensing weighting factor $ \varpi_s $.  Despite the attenuation in linear power, the nonlinear power at the user direction of the proposed algorithm has a remarkable $ 42 \,$dB reduction. The notch of the nonlinear pattern also appears at $ 60^\circ $ where the target is located. This is because the proposed algorithm tends to suppress the nonlinear distortion in the beamformed directions. 
\section{Conclusion}\label{V}
In this paper, we {analyze the performance of a partially-connected hybrid beamforming ISAC system with nonlinear amplification. 
{For this system,} we propose a distortion-aware beamforming algorithm to maximize the weighted sum of the communication rate and sensing MI while meeting the overall transmit power constraint. To solve the highly coupled optimization problem, we first solve for a full digital precoder based on the AO and MO algorithms and then obtain the analog and digital beamforming matrices through a decomposition procedure.  As indicated by the numerical results, the stronger the nonlinearity of PAs, the more advantageous the proposed algorithm becomes when compared to the conventional beamforming   schemes.
\appendix
The complex gradient w.r.t. $ \mathbf{F}^* $ in $\mathcal{P}3$ is given by:
\begin{eqnarray}
	\begin{aligned}
	&\nabla_{\mathbf{F}^*}\tilde{\mathcal{G}}_1=\varpi_c\sum_{k=1}^K\frac{\log_2(e)}{(1+S_k/N_k)}\frac{\left(N_k\frac{\partial S_k}{\partial\mathbf{F}^*}-S_k\frac{\partial N_k}{\partial\mathbf{F}^*}\right)}{N_k^2}+\\&\varpi_s\frac{\log_2(e)}{(1+S_s/N_s)}\frac{\left(N_s\frac{\partial S_s}{\partial\mathbf{F}^*}-S_s\frac{\partial N_s}{\partial\mathbf{F}^*}\right)}{N_s^2}+\lambda_1\frac{\partial \mathcal{C}_1}{\partial\mathbf{F}^*}+\lambda_2\frac{\partial \mathcal{C}_2}{\partial\mathbf{F}^*},
	\end{aligned}
\end{eqnarray}
where we define $ N_k\triangleq I_k+D_k+\sigma^2_k $, $ N_s\triangleq D_s+\sigma^2_s $, $ 	\mathcal{C}_1\triangleq 		\left\| \mathbf{U}-\mathbf{F}\mathbf{F}^H\odot\mathbf{F}^*\mathbf{F}^T\right\|_F^2 $, $ 	\mathcal{C}_2\triangleq \left\| \mathbf{V}-\mathbf{U}\odot\mathbf{F}\mathbf{F}^H\right\|_F^2 $.
Starting with $ S_k $, according to the matrix calculus,
%
	\begin{equation}
	\begin{aligned}
		&\frac{\partial S_k}{\partial\mathbf{F}^*}=2\beta_3\text{Diag}\left(\mathbf{f}_k\odot\mathbf{h}_k^* \right)\mathbf{F}\left( \mathbf{f}^H_k\mathbf{B}^H\mathbf{h}_k\right) +\mathbf{h}^H_k\mathbf{B}\mathbf{f}_k\big(\beta_1^* \mathbf{h}_k\mathbf{i}^T_k\\&+2\beta_3^*\text{Diag}\left(\mathbf{f}_k^*\odot\mathbf{h}_k \right)\mathbf{F}+2\beta_3^*\mathbf{h}_k\odot\mathrm{diag}\left( \mathbf{I}_{N_t}\odot\mathbf{F}\mathbf{F}^H\right) \mathbf{i}_{k}^T \Big),
	\end{aligned}
\end{equation}%
where $ \mathbf{i}_k $ is the $ k $-th standard basis vector.  Similarly,
	\begin{equation}
	\begin{aligned}
		&\frac{\partial I_k}{\partial\mathbf{F}^*}=\sum_{i\neq k}2\beta_3\text{Diag}\left(\mathbf{f}_i\odot\mathbf{h}_k^* \right)\mathbf{F}\left( \mathbf{f}^H_i\mathbf{B}^H\mathbf{h}_k\right) +\mathbf{h}^H_k\mathbf{B}\mathbf{f}_i\big(\beta_1^* \mathbf{h}_k\mathbf{i}^T_i\\&+2\beta_3^*\text{Diag}\left(\mathbf{f}_i^*\odot\mathbf{h}_k \right)\mathbf{F}+2\beta_3^*\mathbf{h}_k\odot\mathrm{diag}\left( \mathbf{I}_{N_t}\odot\mathbf{F}\mathbf{F}^H\right) \mathbf{i}_{i}^T \Big),
	\end{aligned}
\end{equation} %
	\begin{equation}
	\begin{aligned}
		&\frac{\partial S_s}{\partial\mathbf{F}^*}=\left|\alpha_s \right|^2 \sum_i2\beta_3\text{Diag}\left(\mathbf{f}_i\odot\mathbf{a}_s^* \right)\mathbf{F}\left( \mathbf{f}^H_i\mathbf{B}^H\mathbf{a}_s\right) +\mathbf{a}^H_s\mathbf{B}\mathbf{f}_i\big(\beta_1^* \mathbf{a}_s\mathbf{i}^T_i\\&+2\beta_3^*\text{Diag}\left(\mathbf{f}_i^*\odot\mathbf{a}_s \right)\mathbf{F}+2\beta_3^*\mathbf{a}_s\odot\mathrm{diag}\left( \mathbf{I}_{N_t}\odot\mathbf{F}\mathbf{F}^H\right) \mathbf{i}_{i}^T \Big).
	\end{aligned}
\end{equation} %
Furthermore, with $\mathbf{H}_k=\mathbf{h}_k\mathbf{h}_k^H $ and 
$ \mathbf{A}_s=\mathbf{a}_s\mathbf{a}_s^H $, we have
	\begin{eqnarray}\frac{\partial D_k}{\partial\mathbf{F}^*}=2|\beta_3|^2 \left(2\mathbf{H}_k\odot\mathbf{C_x}\odot\mathbf{C_x}^*+\mathbf{H}^*_k\odot\mathbf{C_x}\odot\mathbf{C_x}\right)\mathbf{F}, \end{eqnarray} 
	\begin{equation}\frac{\partial D_s}{\partial\mathbf{F}^*}=2|\beta_3|^2|\alpha_s|^2 \left(2\mathbf{A}_s\odot\mathbf{C_x}\odot\mathbf{C_x}^*+\mathbf{A}^*_s\odot\mathbf{C_x}\odot\mathbf{C_x}\right)\mathbf{F}, \end{equation} 
	\begin{equation}\frac{\partial C_1}{\partial\mathbf{F}^*}=4 \mathbf{C_x}\odot\mathbf{C_x}\odot\mathbf{C_x}^*\mathbf{F} -2\left(\Re\left\lbrace\mathbf{U} \right\rbrace +\Re\left\lbrace\mathbf{U}^T \right\rbrace 	\right)\odot\mathbf{C_x}\mathbf{F}, \end{equation} 
	\begin{equation}%
		\begin{aligned}
\frac{\partial C_2}{\partial\mathbf{F}^*}&=\left(  \mathbf{U}\odot\mathbf{U}^*\odot\mathbf{C_x}+\mathbf{U}^T\odot\mathbf{U}^H\odot\mathbf{C_x}\right) \mathbf{F}\\&-\left( \mathbf{V}\odot\mathbf{U}^*+\mathbf{V}^H\odot\mathbf{U}^T\right) \mathbf{F}.
	\end{aligned}
 \end{equation} %

\end{document}